\documentclass[12pt]{elsart}
\usepackage{graphicx,epsfig}

\newcommand{\be} {\begin{eqnarray*}}
\newcommand{\ee} {\end{eqnarray*}}
\newcommand{\bi} {\begin{itemize}}
\newcommand{\ei} {\end{itemize}}

\newcommand{\bcen}{\begin{center}}
\newcommand{\ecen}{\end{center}}
\newcommand{\beq}{\begin{equation}}
\newcommand{\eeq}{\end{equation}}
\newcommand{\bea}{\begin{eqnarray}}
\newcommand{\eea}{\end{eqnarray}}
\newcommand{\ba}{\begin{array}}
\newcommand{\ea}{\end{array}}
\newcommand{\bann}{\begin{eqnarray*}}
\newcommand{\eann}{\end{eqnarray*}}

\begin{document}
\begin{frontmatter}
\title{Nonlocality effects on spin-one pairing patterns
in two-flavor color superconducting quark matter and 
compact star applications
}

\author[rostock]{D. N. Aguilera}
and
\author[dubna,gsi]{D. B. Blaschke}
\address[rostock]{Institut f\"ur Physik, Universit\"at Rostock, \\
Universit\"atsplatz 3, D-18051 Rostock, Germany}
\address[dubna]{Bogoliubov  Laboratory of Theoretical Physics, JINR Dubna,\\
Joliot-Curie street 6, 141980  Dubna, Russia}
\address[gsi]{Gesellschaft f\"ur Schwerionenforschung (GSI), \\
Planckstr. 1, 64291 Darmstadt, Germany}

\begin{abstract}
We study the influence of nonlocality in the interaction on two
spin one pairing patterns of two-flavor quark matter: 
the anisotropic blue color paring besides the usual
two color superconducting matter (2SCb), in which red and green colors are 
paired, and the color spin locking phase (CSL).
The effect of nonlocality on the gaps is rather large and
the pairings exhibit a strong dependence on the form factor of the interaction,
especially in the low density region.
The application of these small spin-one condensates for compact stars is 
analyzed: the early onset of quark matter in the nonlocal models may help to 
stabilize hybrid star configurations.
While the anisotropic blue quark pairing does not survive a big asymmetry in 
flavor space as imposed by the charge neutrality  condition, 
the CSL phase as a flavor independent pairing can be realized as
neutral matter in compact star cores. 
However, smooth form factors and the missmatch between the flavor chemical 
potential  in neutral matter make the effective gaps of the order of magnitude 
$\simeq 10$ keV, and a more systematic analysis 
is needed to decide whether such small gaps could be consistent with 
the cooling phenomenology.

\noindent PACS number(s):  04.40.Dg, 12.38.Mh, 26.60.+c, 97.60.Jd
\end{abstract}
\end{frontmatter}
\newpage
\section{Introduction}

The investigation of color superconducting phases in cold dense quark matter 
has attracted much interest in the last years since those phases could be 
relevant for the physics of compact star cores \cite{Blaschke:2001uj}. 
Nevertheless, it became clear that large spin-0 condensates, like the usual 
two-flavor superconductor (2SC), although having large pairing gaps 
($\sim$ 100 MeV) and therefore a direct influence on the equation of state 
(EoS), may be disfavored by the charge neutrality condition if not unusually 
strong diquark coupling constants are considered 
\cite{Aguilera:2004ag,Blaschke:2005uj,Ruster:2005jc}. 

Alternatively, spin-1 condensates are being investigated 
\cite{Schafer:2000tw,Alford:2002rz,Schmitt:2004et}.
Due to their smallness, their influence on the EoS is negligible but they 
could strongly  affect the transport properties in quark matter and therefore 
have important consequences on observable phenomena like compact star cooling, 
see \cite{Blaschke:2005dc}. 
A recent investigation of neutrino emission and cooling rates of spin-1 color 
superconductors constructed for conserved total angular momentum allows for 
color-spin locking, planar, polar, and {\it A} phases \cite{Schmitt:2005wg}. 
However,  none of these phases fulfills the requirements of cooling 
phenomenology that no ungapped quark mode should occur on which the direct 
Urca process could operate and lead to very fast cooling 
in disagreement with modern observational data \cite{Grigorian:2004jq}.  

In the present work, we consider spin-1 pairing patterns different from 
the above mentioned, like  the anisotropic third color (say blue) quark 
pairing besides the usual 2SC phase (2SCb) \cite{Buballa:2002wy} and the 
s-wave color-spin locking phase (CSL) \cite{Aguilera:2005tg}, which  have been 
introduced within the NJL model whereby small gaps in the region of  some 
fractions of MeV have been obtained.
Such small gaps could help to suppress efficiently the direct Urca process in 
quark matter and thus to control the otherwise too rapid cooling of hybrid 
stars \cite{Grigorian:2004jq,Popov:2005xa}.

One important feature is that the form of the regularization, i.e. via a sharp 
cut-off or a form factor function, is expected to have a strong impact on the 
resulting gaps due to the sensitive momentum dependence of the integrand
in the gap equation \cite{Buballa:2002wy}.
Especially the behavior of quark matter in the density region of the suspected 
deconfinement transition plays a crucial role for determining the stability of 
compact star configurations. 
Models with a late onset of quark matter could  eventually lead to unstable 
hybrid star configurations. 

For example, in \cite{Aguilera:2005tg} it has been shown within the local NJL 
model how a possible pairing pattern for compact star matter could be
described which fulfills the constraints from compact star cooling 
phenomenology. 
These require that all quark species be paired and the smallest pairing gap be 
of the order of 10 keV to 1 MeV with a decreasing density dependence.
A caveat of the NJL model quark matter is, however, that a stable hybrid star 
can be realized only marginally, see \cite{Baldo:2002ju}.

The advantage of nonlocal models is that they can describe
the regularization of the quark interaction via form factor functions and 
therefore represent it in a smoother way, especially for low densities
\cite{Schmidt:1994di}.
For  the case of the 2SC phase it has already been shown that the effect of 
nonlocality in the low density region is rather large
and the pairing exhibits a strong dependence  with the form factor of the 
interaction \cite{Aguilera:2004ag,Blaschke:2003yn}.
Moreover, the early onset of quark matter for the dynamical chiral quark model,
in contrast to the NJL model,  might help to stabilize hybrid star 
configurations.

As it has been shown in \cite{Grigorian:2003vi} within a nonlocal 
generalization of the NJL model \cite{Schmidt:1994di}, stable hybrid stars 
with large quark matter cores can be obtained.
In order to describe the properties of these stars consistently, including 
their cooling phenomenology, the description of diquark pairing gaps as well 
as emissivities and transport properties for the above motivated CSL phase 
should be given also within a nonlocal quark model.

As a first step in this direction we will provide in the present 
paper the spin-1 pairing gaps for a nonlocal, instantaneous chiral quark 
model under neutron star constraints for later application in the 
cooling phenomenology.
We give here a first discussion of the influence of nonlocality of
the interaction in momentum space when compared to the NJL model case and 
discuss the possible role of spin-one condensates for compact star 
applications.

The paper is organized as follows: in Section 2 we briefly review how the NJL 
model is modified when nonlocality is introduced using a 
three-\-dimensional nonlocal chiral quark model; in Sections 3  and 4  
we present the nonlocal version of the anisotropic blue color paring (2SCb) 
besides the usual two-flavor color superconducting (2SC) phase and of the 
color-spin locking phase (CSL), respectively. 
In Section 5 we present preliminary results for neutral matter in compact 
stars for the Gaussian form factor and discuss whether requirements for hybrid 
star cooling phenomenology could be met.  
Finally, in Section 6,  we draw the Conclusions.


\section{Nonlocal chiral quark model}

We investigate a nonlocal chiral model for two-flavor quark matter in which
the quark interaction is represented in a separable way by introducing 
form factor functions $g(p)$ in the bilinears of the current-current 
interaction terms in the Lagrangian 
\cite{Aguilera:2004ag,Schmidt:1994di,Blaschke:2003yn,Grigorian:2003vi}.
It is assumed that this four-fermion interaction is instantaneous
and therefore the form factors depend only on the modulus of the
three momentum $p=|\vec p|$.

In the mean field approximation the thermodynamical potential can be evaluated 
and is given by
\bea
\Omega(T,\mu)&=&-T \sum_{n} \int \frac{d^3p}{(2\pi)^3}
\frac{1}{2}{\rm Tr}\ln({\frac{1}{T}S^{-1}(i\omega_n,\vec p\,)}) + V~,
\label{Omega}
\eea
where the sum is over fermionic Matsubara frequencies
$\omega_n = (2n+1)\pi T$ and  $V$ is the quadratic contribution of the 
condensates considered.
The specific form of $V$ in dependence on the order parameters $\phi$ for 
chiral symmetry breaking and $\Delta$ for color superconductivity in the 
corresponding diquark pairing channels will be given below in Section 3.

In our nonlocal extension the inverse of the fermion propagator in 
Nambu-Gorkov space is modified in comparison to the NJL model case by momentum 
dependent form factor functions $g(p)$ as follows,
\beq
S^{-1}(p)=
\left(
\begin{array}{cc}
 \not\!p +\hat \mu\gamma^0-\hat M(p)&
g(p)\hat\Delta\\
-{g(p)\hat\Delta}^\dagger&
\not\!p -\hat \mu\gamma^0-\hat M(p)
\end{array}
\right)
\label{InvProp}
\eeq
where  $\hat{\mu}$ is the chemical potential matrix and the elements of
$\hat{M}(p)=\mathrm{diag}\{M_f(p)\}$ are
the dynamical masses of the quarks given by
\bea
M_f(p)=m_f+g(p)\phi_f~.
\label{Mass}
\eea
The matrix $\hat \Delta$ represents the order parameters for diquark pairing 
which will be made explicit in Sects. 3 and 4, respectively.

In  (\ref{InvProp}) and (\ref{Mass}) we have introduced the same form factors 
$g(p)$ to represent the nonlocality of the interaction in the meson 
($q \bar q$) and diquark ($qq$) channels.
In our calculations we use the Gaussian (G), Lorentzian (L) and cutoff (NJL)
form factors defined as
\begin{eqnarray}
\label{GF}
g_{\rm G}(p) &=& \exp(-p^2/\Lambda_{\rm G}^2)~,\\
\label{LF}
g_{\rm L}(p) &=& [1 + (p^2/\Lambda_{\rm L}^2)^2]^{-1},\\
\label{NF}
g_{\rm NJL}(p) &=& \theta(1 - p/\Lambda_{\rm NJL})~.
\end{eqnarray}
The parameters for the above form factor models used in this work 
are  presented in Tab. \ref{parNJL}. 
They have been fixed by the pion mass, pion decay constant and the constituent 
quark mass $M=M(0)$ at $T=\mu=0$. 
In order to estimate the effect of the nonlocality on the results relative to 
the NJL model,  we used $g_G$, $g_L$ and $g_{NJL}$ with parameters fixed such
that $M=380$ MeV, see Tab. \ref{parNJL}.
For details of the parameterization, see \cite{Grigorian:2006qe}.

\begin{table}[htb]
\begin{center}
\begin{tabular}{|l|c||c|c|c|}\hline
Form Factor  & Notation &$\Lambda$[MeV] &$G~\Lambda^2$ &$m$[MeV]\\ 
\hline
Gaussian   &$g_G$    &$786.7$  &$4.12$&$2.50$\\
Lorentzian  &$g_L$    &$637.2$  &$2.76$&$2.59$\\
NJL         &$g_{NJL}$&$596.1$ &$2.36$&$5.54 $\\
\hline
  \end{tabular}
\vspace{1cm}
\caption{Parameter sets for the nonlocal chiral quark model ($g_G$, $g_L$ ) 
and for the NJL model ($g_{NJL}$); for all $M(0)=380$ MeV is fixed.  }
     \vspace*{0.5cm}
  \label{parNJL}
\end{center}
\end{table}

The stationary points of the  thermodynamical potential (\ref{Omega})
are found from the condition of a vanishing variation 
\bea
\delta \Omega =0~
\label{dOmega}
\eea
with respect to variations of the order parameters.
Eq. (\ref{dOmega}) defines a set of gap equations.
Among the solutions of these equations the thermodynamically stable state 
corresponds to the set of order parameter values for which $\Omega$ has an 
absolute minimum.

\section{Anisotropic blue quark pairing for nonlocal model}

First we consider the 2SCb phase in which
two of the three colors (e.g. red $r$ and green $g$)
pair in the standard spin-0 isospin singlet condensate (2SC phase) and the 
residual third color (consequently, it is blue $b$) pairs in a spin-1 
condensate (symmetric in Dirac space, symmetric in color, antisymmetric in 
flavor) \cite{Buballa:2002wy}.
The matrix $\hat \Delta$ in the inverse quark propagator (\ref{InvProp}) for 
the 2SCb  phase is then given by
\bea
\hat\Delta^{2SCb}=\Delta(\gamma_5\tau_2\lambda_2)(\delta_{c,r}+\delta_{c,g})
+\Delta^{\prime}(\sigma^{03}~\tau_2~\hat P_3^{(c)})\delta_{c,b}~,
\eea
where $\tau_2$ is an antisymmetric Pauli matrix in the flavor space 
and $\lambda_2$ an antisymmetric Gell-Mann matrix in the color space;
$\sigma^{03}=\frac i2 [\gamma^0,\gamma^3]$ and 
$\hat P_3^{(3)} = \frac13-\frac{\lambda_8}{\sqrt{3}}$ is the projector on the 
third color. 
If  $\Delta^{\prime} \ne 0$, this can be understood as a nonzero third 
component of a vector in color space which breaks the $\mathcal{O}(3)$ 
rotational symmetry spontaneously.
The blue quark pairing is therefore anisotropic. 
We consider first symmetric matter and thus we take the quark chemical 
potentials as $\mu_u=\mu_d=\mu$.  

The thermodynamical potential is given by
\bea
\Omega^{2SCb}(T,\mu)&=&
\frac{\phi^2}{4G_1}+\frac{|\Delta|^2}{4G_2}
+\frac{|\Delta^{\prime}|^2}{16G_3}\nonumber
\\
-
4 \sum_{i=1}^{3} \int &&\frac{d^3p}{(2\pi)^3}
\left[
\frac{E_i^-+E_i^+}{2}
+T \ln{(1+e^{-E_i^-/T})}+T \ln{(1+e^{-E_i^+/T})}
\right]~,\nonumber\\
\label{Omegablue}
\eea
where the coupling constants $G_1,G_2,G_3$ follow the relation given by the 
instanton induced interaction \cite{Buballa:2003qv}
\bea
G_1:G_2:G_3=1:3/4:3/16~.
\eea
The dispersion law for the paired quarks ($r\, , \,g$) is given by
\bea
E_{1,2}^{\mp}(\vec p) &=& 
E^{\mp}(\vec p)=\sqrt{(\epsilon \mp \mu)^2 +g^2(p)|\Delta|^2 }~,
\label{E_12}
\eea
where $\epsilon= \sqrt{\vec p\,^2+M^2}$ is the free particle dispersion 
relation and $M=M_u=M_d$.

For the anisotropic pairing of the blue quarks the dispersion relation can be 
written as
\bea
E_{3}^{\mp}(\vec p) &=& \sqrt{(\epsilon_{\rm eff}\mp \mu_{\rm eff})^2 
+g^2(p)|\Delta^{\prime}_{\rm eff}|^2 }~,
\label{E_3}
\eea
where the effective variables depend  on the angle $\theta$, 
with $\cos \theta= p_3/|\vec p|$, and are defined as
\bea
\epsilon_{\rm eff}^2&=& \vec p\,^2+M_{\rm eff}^2~,\\
M_{\rm eff}&=&M\frac{\mu}{\mu_{\rm eff}}~,\\
\mu_{\rm eff}^2&=&\mu^2+g^2(p)|\Delta^{\prime}|^2\sin^2{\theta}~,\\
|\Delta^{\prime}_{\rm eff}|^2 &=& |\Delta^{\prime}|^2(\cos^2\theta 
+ \frac{M^2}{\mu_{\rm eff}^2}\sin^2{\theta}).
\label{effective}
\eea

The dispersion relation $E_3^{\pm}$ is, therefore, an anisotropic function of 
$\vec p$ and therefore the calculation of (\ref{Omegablue}) should be 
performed as an integral over the modulus of $|\vec p|$ and over the angle 
$\theta$. 
$E_3^{\pm}$ has a  minimum if $\theta=\pi/2$ and vanishes if $M=0$ or 
$\Delta^{\prime}=0$.

\begin{figure}[bth]
  \begin{center}
    \includegraphics[width=0.75\linewidth,angle = -90]{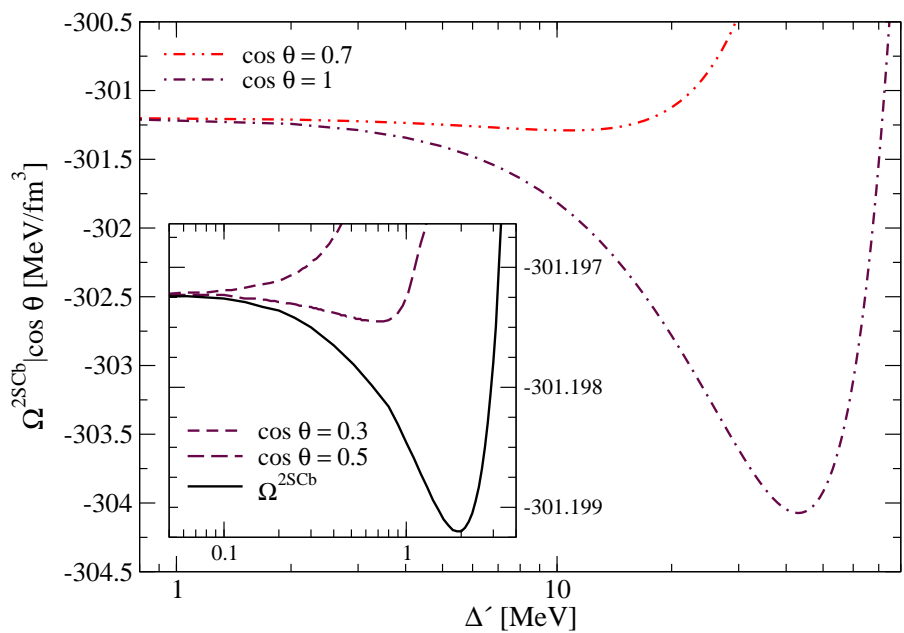}
     \vspace*{-0.5cm}
    \caption{Anisotropic contributions $\Omega^{2SCb}\,|_{\cos\theta}$ 
to the thermodynamical potential  $\Omega^{2SCb}$
as a function of the spin-1 gap $\Delta^{\prime}$ for fixed 
$(\mu=500$ MeV, T=0) and ($\phi=20$ MeV, $\Delta=106$ MeV).
As $\theta$ increases the position of the minimum of 
$\Omega^{2SCb}\,|_{\cos\theta}$  moves to lower values of the gaps 
$\Delta^{\prime}$.
The low gap region is zoomed in the inset figure on the bottom left.
For $\Omega^{2SCb}$ (solid line) we obtain that the minimum is placed at 
$\Delta^{\prime}\simeq2$ MeV/fm$^3$. The NJL parameterization is considered.}
    \label{Omegacos}
  \end{center}
\end{figure}

As it has been pointed out in \cite{Buballa:2002wy}, the gap equations for 
$\Delta$ and $\Delta^{\prime}$ are only indirectly coupled by their dependence 
on $M$. 
Therefore, since the equation for $\Delta^{\prime}$ nearly decouples if $M$ is 
small, we can illustrate the anisotropic contributions to the thermodynamical 
potential $\Omega$ fixing the variables $(\mu,T)$ and the order parameters 
$(\phi,\Delta)$ and varying the angle $\theta$.
For this purpose, we consider
\bea
d\Omega^{2SCb}= d(\cos \theta) \Omega^{2SCb}\,|_{\cos\theta}~,
\eea
and in Fig. \ref{Omegacos} we plot $\Omega^{2SCb}\,|_{\cos\theta}$ as a 
function of the gap $\Delta^{\prime}$.   
As $\theta$ increases from $0$ to $\pi/2$ the position of the minimum of 
$\Omega^{2SCb}\,|_{\cos\theta}$ moves to lower values of the gaps 
$\Delta^{\prime}$. 
The value of $\Delta^{\prime}$ that minimizes the thermodynamical potential 
is found once the integration over the angle $\theta$ is performed. 

\subsection{Gap equation solutions}

We search for the stationary points of $\Omega^{2SCb}$ (\ref{dOmega}) respect 
to the order parameters solving the gap equations
\bea
\frac{\delta\Omega^{2SCb}}{\delta\phi}=\frac{\delta\Omega^{2SCb}}{\delta\Delta}
=\frac{\delta\Omega^{2SCb}}{\delta\Delta^{\prime}}=0
\label{GEblue}
\eea
using the dispersion relations (\ref{E_12}) and (\ref{E_3}). 
The results that are shown in this work are obtained for $T=0$. 

\begin{figure}[bth]
  \begin{center}
    \includegraphics[width=0.9\linewidth,height=\linewidth, angle = -90]
{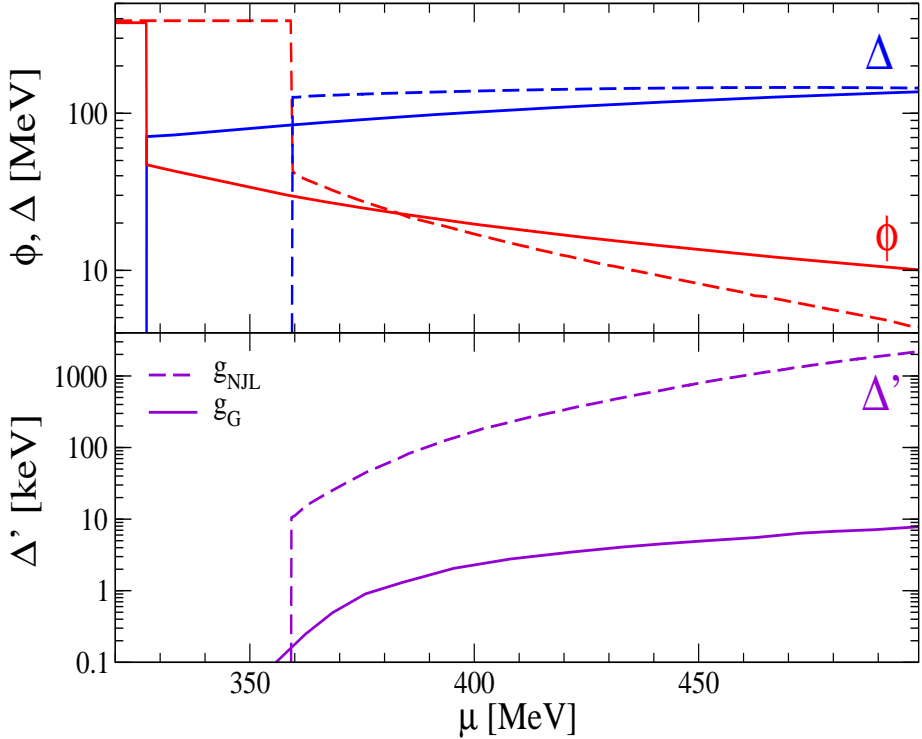}
    \caption{Chiral gap $\phi$,   2SC diquark gap $\Delta$ (upper panel) and 
spin-1 pairing gap of the blue quarks $\Delta^{\prime}$ (lower panel) as a 
function of the quark chemical potential $\mu$ for the NJL model and for the 
Gaussian form factor. 
}
    \label{GE_NJL_blue}
  \end{center}
\end{figure}

In Fig. \ref{GE_NJL_blue} we show the chiral gap $\phi$, 
the 2SC diquark gap $\Delta$ and the spin-1 pairing gap of the blue quarks
$\Delta^{\prime}$  as functions of the quark chemical potential $\mu$. 
The gaps $\Delta^{\prime}$   are strongly density-dependent rising functions 
and typically of the order of magnitude of keV, e.g., for a fixed $\mu$ at 
least two orders of magnitude smaller than the corresponding 2SC gaps.
These small gaps are very sensitive to the form of the regularization and to 
the parameterization used. 
Obviously, also the onset and the slope of the superconducting phases
depend strongly on the parameters used.

The effect of the nonlocality  on the results is shown in the lower panel of 
Fig. \ref{GE_NJL_blue}: the smoothness of the Gaussian form factor reduces the 
$\Delta^{\prime}$ gaps dramatically. 
For this case, the blue quark pairing gaps are about two orders of magnitude 
lower than the corresponding NJL model (both with fixed $M=380$ MeV).  

\begin{figure}[bth]
  \begin{center}
    \includegraphics[width=0.7\linewidth,height=\linewidth,angle = -90]
{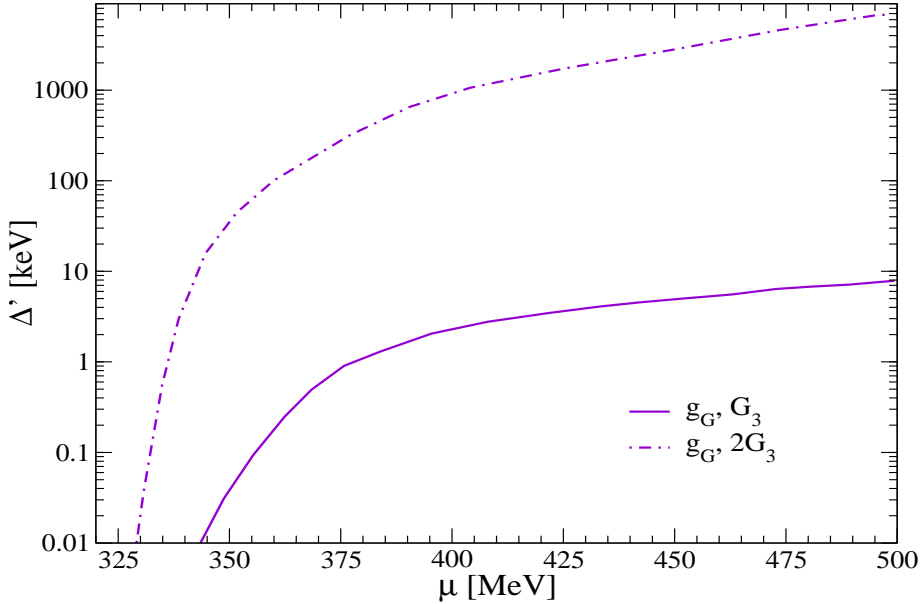}
    \caption{Effect of the coupling constant $G_3$ on the spin-1 blue quarks 
energy gaps $\Delta^{\prime}$ for the Gaussian form factor. 
The dash-dotted line corresponds to the case when the coupling constant $G_3$ 
is doubled.}
    \label{GE_FF_blue}
  \end{center}
\end{figure}

The dependence of the results on the parameters used is also rather strong and 
nonlinear as it is shown in Fig.~\ref{GE_FF_blue} using, as example, the 
Gaussian form factor.  
When the coupling constant $G_3$ is doubled (dash-dotted line) the resulting 
gaps increase between two and three orders of magnitude, depending on the 
chemical potential. 

Since these small gaps are practically negligible in comparison to usual 2SC 
gaps, they would have no influence on the equation of state. 
On the other hand, it is well known that even small pairing energies could play
an important role in the calculation of transport properties of quark matter
for temperatures below the order of magnitude of the gap parameter.
Nevertheless, it is unlikely that the blue quark pairing could survive the 
compact star constraints: the charge neutrality requires that the Fermi seas 
of the up and down quarks should differ by about 50-100 MeV and this is much 
larger than the gaps that we obtain for these condensates ($\sim$ keV) in the 
symmetric case.

In the following section we present the nonlocality effects on a flavor 
symmetric spin-1 pairing channel being a good candidate to survive the large 
mismatch between up and down quark Fermi seas in charge neutral quark matter.

\section{Color-spin locking (CSL) phase for a nonlocal chiral model}

In the s-wave CSL phase introduced in Ref. \cite{Aguilera:2005tg}, which 
differs from the CSL phase in \cite{Schmitt:2004et}, each condensate is a 
component of the antisymmetric anti-triplet in the color space and is locked 
with a vector component in the spin space.  
In the present paper, we study a nonlocal generalization of the CSL pairing 
pattern of Ref. \cite{Aguilera:2005tg} and consider the matrix $\hat \Delta$ 
in (\ref{InvProp}) for the CSL channel as
\bea
\hat\Delta^{CSL}=
\Delta_f(\gamma_3\lambda_2+\gamma_1\lambda_7+\gamma_2\lambda_5)~.
\label{hatD_csl}
\eea
The thermodynamical potential can be decomposed into single-flavor components
\beq
\Omega^{CSL}(T,\{\mu_f\}) = \sum_{f \in \{u,d\}} \Omega^{CSL}_f(T,\mu_f)~,
\label{omegaq}
\eeq
where the contribution of each flavor is
\bea
\Omega^{CSL}_f(T,\mu_f)&=&
 \frac{\phi_f^2}{8G_1}
+3\frac{|\Delta_f|^2}{8H_v}
- \sum_{k=1}^6 \int \frac{d^3p}{(2\pi)^3}
(E_{f,k}+2T\ln{(1+e^{-E_{f,k}/T})})~. \nonumber\\
\eea
The ratio of the two coupling constants
\bea
G_1:H_v=1:\frac38.
\eea
is obtained via Fierz transformations of the color-currents
for a one-gluon exchange interaction.
To derive the dispersion relations $E_{f,k}$ we follow \cite{Aguilera:2005tg} 
and we extend the  expressions for the nonlocal model introducing the form 
factors to modify the quark interaction.
We obtain that $E_{f;1,2}$ could be brought in the standard form
\bea
E_{f;1,2}^2&=& (\varepsilon_{f,{\rm eff}}\mp\mu_{f,{\rm eff}})^2+
|\Delta_{f, {\rm eff}}|^2g^2(p)
\eea
if the effective variables are now defined as:
\bea
\varepsilon_{f, {\rm eff}}^2&=& \vec p\,^2+M_{f,\rm eff}^2~,\\
M_{f,\rm eff} &=& \frac{\mu_f}{\mu_{f,{\rm eff}}}M_f(p)~, \\
\mu_{f,{\rm eff}}^2&=&\mu_f^2+|\Delta_f|^2 g^2(p)~,\\
\Delta_{f,\rm eff}&=& \frac{M_f(p)}{\mu_{f,{\rm eff}}} |\Delta_f|~.
\label{effective2}
\eea

For $E_{f;k}$, $k=3...6$, the particle and the antiparticle branches split
\bea
E_{f;3,5}^2&=& (\varepsilon_{f}-\mu_{f})^2+
a_{f;3,5}|\Delta_f|^2 g^2(p)~,
\\
E_{f;4,6}^2&=& (\varepsilon_{f}+\mu_{f})^2+
a_{f;4,6}|\Delta_f|^2 g^2(p)~,
\eea
where the momentum-dependent coefficients $a_{f;k}$, $k=3...6$ are given by
\bea
a_{f;3,5}&=&\frac{1}{2}\left[5-\frac{\vec p\,^2}{\varepsilon_f\mu_f}
\pm \sqrt{\left(1-\frac{\vec p\,^2}{\varepsilon_f\mu_f}\right)^2
+8\frac{M_f^2(p)}{\varepsilon_f^2}}
\right] \\
a_{f;4,6}&=&\frac{1}{2}\left[5+\frac{\vec p\,^2}{\varepsilon_f\mu_f}
\pm \sqrt{\left(1+\frac{\vec p\,^2}{\varepsilon_f\mu_f}\right)^2
+8\frac{M_f^2(p)}{\varepsilon_f^2}}
\right]
\label{coeffs}
\eea
and
\bea
\varepsilon_f^2&=& \vec p\,^2+M_f^2(p)~.
\label{varepsilon}
\eea

We solve the gap equations
\bea
\frac{\delta\Omega^{CSL}_f}{\delta\phi_f}=
\frac{\delta\Omega^{CSL}_f}{\delta\Delta_f}=0
\label{GE_csl}
\eea
and present the results of the global minimum of $\Omega^{CSL}_f$ in the next 
subsection.

\subsection{Gap equation solutions for each flavor}

\begin{figure}[bth]
  \begin{center}
    \psfig{figure=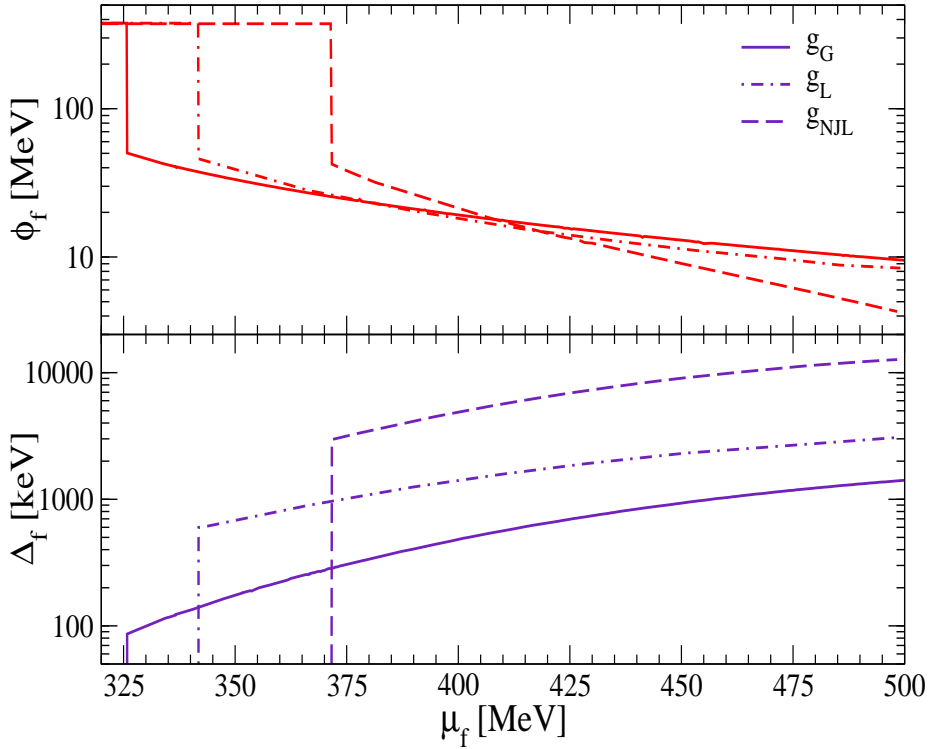,width=0.9\textwidth,height=\linewidth,angle=-90}
     \vspace*{0.5cm}
    \caption{The dependence of the chiral gap $\phi$ and the CSL pairing gap 
     $\Delta_f$ on the chemical potential $\mu_f$ for different form factors
    of the nonlocal interaction and for the NJL model.}
    \label{FF_gaps_1.1}
  \end{center}
\end{figure}

The  mass gaps and the CSL gaps are shown in Fig. \ref{FF_gaps_1.1} for 
different form factors of the quark interaction. 

The CSL gaps are strongly $\mu_f$-dependent rising
functions in the domain that is relevant for compact star applications.
There is a systematic reduction of the CSL gaps as the form factors 
become smoother (from NJL to Gaussian) and the condensates in the nonlocal 
extension are at least one order of magnitude smaller than in the NJL case. 

Note that especially the low-density region is qualitatively determined by the 
form of the interaction. 
Since the nonlocality also affects the chiral gap, the breakdown of which is 
a prerequisite for the occurrence of color superconducting phases, we observe 
for the nonlocal models an earlier onset of the superconducting quark matter 
($\mu_{f,{\rm crit}} \le 350$ MeV) than in the NJL model cases. 
As it has been pointed out in \cite{Grigorian:2003vi}, the position of the 
onset is crucial to stabilize hybrid star configurations. 
In general, NJL models present a later onset than nonlocal ones 
\cite{Blaschke:2003yn} and might disfavor the occurrence of stable hybrid star 
configurations with a quark matter core \cite{Buballa:2003et}.

\begin{figure}[bth]
 \begin{center}
    \psfig{figure=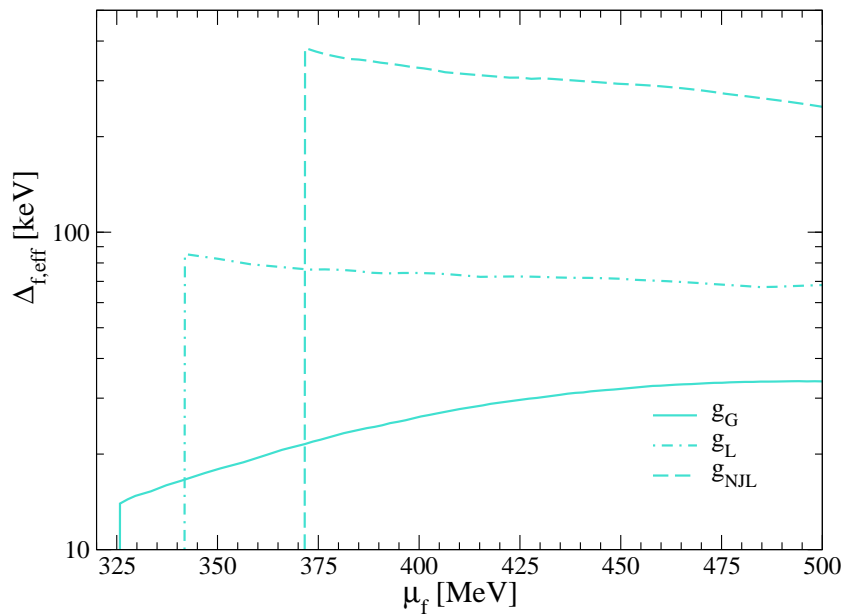,width=0.7\textwidth,angle=-90}
     \vspace*{0.3cm}
    \caption{The effective CSL pairing gap $\Delta_{f,{\rm eff}}$ for 
different form factors as a function of the chemical potential $\mu_f$.}
\label{FF_gaps_3} 
     \vspace*{0.7cm}
  \end{center}
\end{figure}

In  Fig.~\ref{FF_gaps_3}  we plot the effective CSL gaps setting the explicit 
dependence of the form factor $g(p)=1$ in (\ref{effective2}) in order to 
compare the order of magnitude of them with sharp cut-off models. 
We obtain that the Gaussian is  an increasing function of the chemical 
potential from approximately 15 keV to 35 keV and the Lorentzian is nearly 
constant of the order of 80-90 keV.  
Both exhibit gaps that are much smaller than the corresponding NJL ones which 
are in the range $\Delta_{f,{\rm eff}}\simeq$ 300-200 keV.

 \section{Results for matter in compact stars}

Since the CSL pairing is symmetric in flavor, we can easily construct 
electrically neutral quark matter in $\beta$-equilibrium for compact star 
applications.
We consider stellar matter in the quark core of compact stars consisting  
of $\{u,d\}$ quarks and leptons 
$\{e,\nu_e,\bar \nu_e, \mu,\nu_{\mu},\bar \nu_{\mu}\}$.
The particle densities $n_j$ are conjugate to the corresponding
chemical potentials $\mu_j$ according to
\begin{eqnarray}
&&
n_j=-\frac{\partial \Omega}{\partial \mu_j}\bigg|_{\phi_0,\Delta_0;T}~,
\end{eqnarray}
where the index $j$ denotes the particle species.
We consider matter in $\beta$-equilibrium with only electrons
and since we assume that neutrinos leave the star without being trapped 
($\mu_{\bar \nu_e}=-\mu_{\nu_e}=0$, $\mu_{\bar \nu_{\mu}}= -\mu_{\nu_{\mu}}=0$)
\begin{eqnarray}
\mu_e =\mu_d-\mu_u~ .
\label{B_equil}
\end{eqnarray}

We are interested in neutral matter. Therefore we impose that
the total electric charge should vanish
\begin{eqnarray}
\frac{2}{3}n_u-\frac{1}{3}n_d-n_e =0~.
\end{eqnarray}
The CSL condensates are color neutral 
such that no color chemical potentials are
needed and no further constraints need to be obeyed.

We  calculate the CSL gaps for each flavor as a function of  the
quark chemical potential $\mu=(\mu_u+2\mu_d)/3\,$,
\begin{eqnarray}
\Delta_f& =& \Delta_f(\mu_f(\mu))~,\\\nonumber
\Delta_{f,{\rm efff}}& =& \Delta_{f,{\rm eff}}(\mu_f(\mu))~,
\end{eqnarray}
where the functional relation $\mu_f(\mu)$ is taken from the $\beta$-
equilibrated and neutral normal quark matter equation of state.

\begin{figure}[htb]
  \begin{center}
    \includegraphics[width=0.75\linewidth,angle = -90]{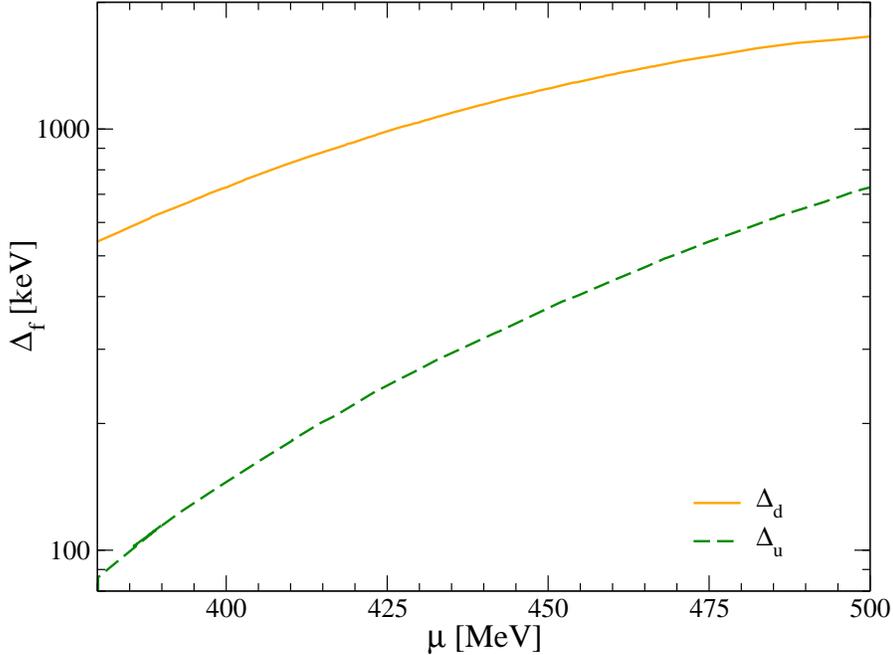}
     \vspace*{0.3cm}
    \caption{CSL gaps for $\beta$-equilibrated and neutral quark matter as a 
    function of the quark chemical potential $\mu$ for the Gaussian set.}
     \vspace*{0.7cm}
    \label{Delta_ud}
  \end{center}
\end{figure}

\begin{figure}[htb]
  \begin{center}
    \includegraphics[width=0.75\linewidth,angle= -90]{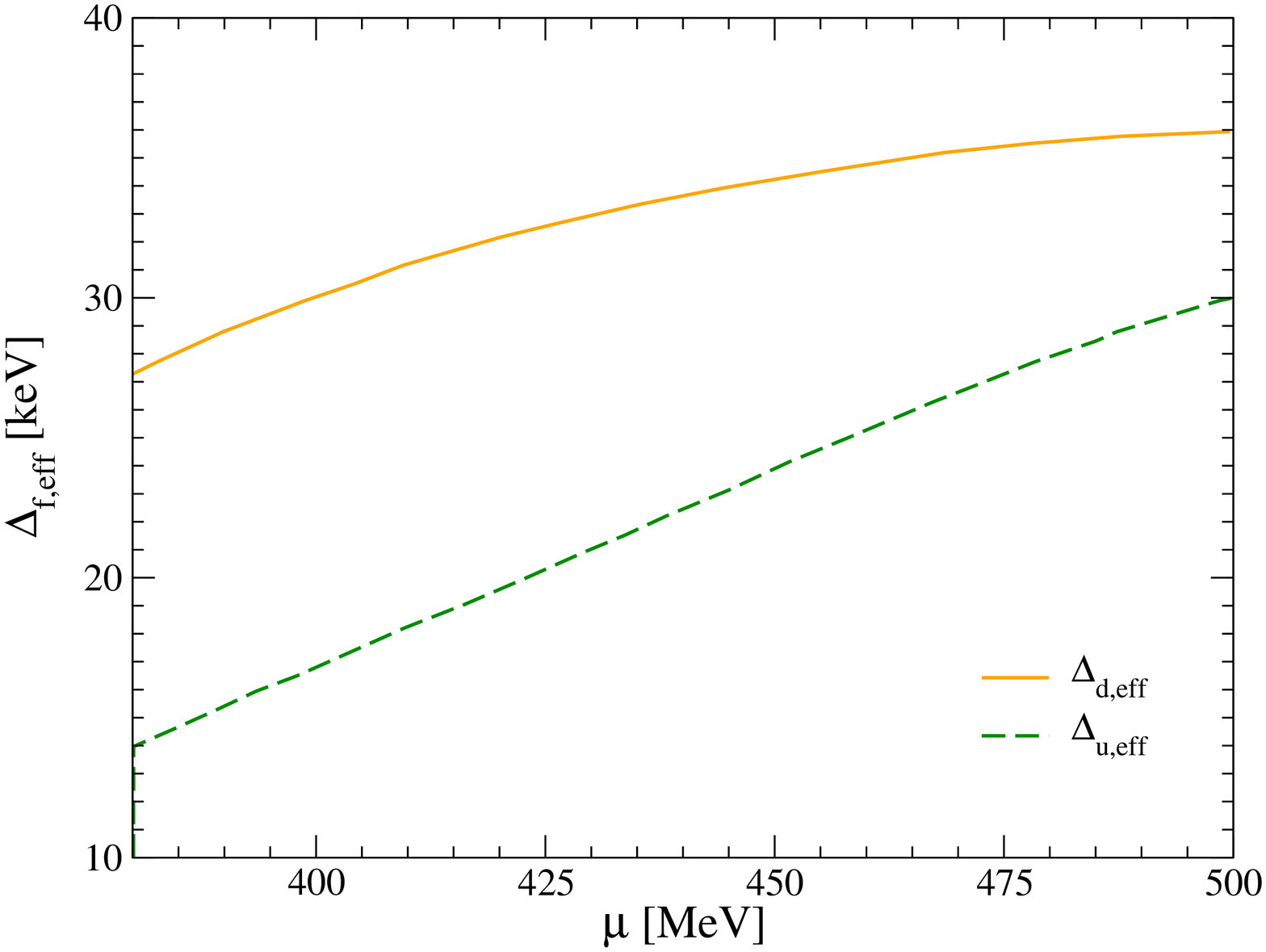}
     \vspace*{0.3cm}
    \caption{Effective CSL gaps for $\beta$-equilibrated and neutral quark 
     matter as a function of $\mu$ for the Gaussian set.}
     \vspace*{0.7cm}
    \label{Delta_eff_ud}
  \end{center}
\end{figure}

The results for the CSL gaps and the effective gaps as a function of the 
chemical potential $\mu$ are shown in Fig. \ref{Delta_ud} and in 
Fig. \ref{Delta_eff_ud}, respectively. 
From the parameterizations we listed in Table 1, we choose the 
Gaussian set because we consider it the most promising for stable hybrid star 
configurations due to the early onset of chiral and superconducting 
phase transitions in quark matter.     

From the figures  \ref{Delta_ud} and \ref{Delta_eff_ud}, we see that the two 
branches of the gap functions corresponding to the up and down quarks are put 
apart by the charge neutrality condition (thick lines). 
The smallest gap, $\Delta_{u}$ runs from $\approx 100$ keV near the onset to 
$\approx 500$ keV at $\mu=500$ MeV while for the $d$ quarks, $\Delta_{d}$ 
increases from $\approx 380$ keV to $1.4$ MeV in the same range. 

On the other hand, the effective gaps $\Delta_{f,{\rm eff}}$ 
are of the order of magnitude of $\simeq 10$ keV, showing an approximate 
linear behaviour with $\mu$. It remains to be investigated whether
such small effective gaps could effectively suppress the direct 
Urca process in quark matter which is a requirement of compact star cooling 
phenomenology.

In this respect, we found here that two facts produce a strong reduction of 
the CSL energy gaps. 
First, when we include smooth form factors in the effective 
interactions, the values of the gaps decrease dramatically relative to the NJL 
case (an order of magnitude from $\simeq 100$ keV for NJL to $\simeq 10$ keV 
for Gaussian). 
Second, when neutrality constraints are considered, $\Delta_{u}$ and the 
effective gaps $\Delta_{f,{\rm eff}}$ are reduced further. 
 
However, as we have shown, there is a strong dependence of our results on the 
parameterization, and a more systematic investigation on the smoothness 
of the form factor could be helpful to 
decide whether these phases could be suitable for compact star applications. 
Moreover, this study should be seen as a preparatory step for subsequent 
investigations where, for example, a covariant generalization of the formalism 
for the inclusion of nonlocality effects 
\cite{Duhau:2004pq,Blaschke:2004cc} should be considered. 

\section{Conclusion}

We have studied the effect of nonlocality on spin-1 condensates in two flavor 
quark matter: the 2SC+spin-1 pairing of the blue quarks (2SCb) and the color 
spin locking (CSL) phase.
We found that the size of these small gaps is very sensitive to the 
form of the regularization. 
The nonlocality has a strong impact on the low density region and we obtain an 
earlier onset for the superconducting phases.
This might be crucial to stabilize quark matter cores in hybrid stars.

On the other hand, due to the flavor asymmetry, we find that the 2SCb pairing 
phase can be ruled out for compact stars applications.
The CSL phase, in contrast, is flavor independent and therefore inert against 
the constraint of electric neutrality. 
For electrically neutral quark matter in beta equilibrium we obtain effective 
CSL gaps which are of the order of magnitude of $10$ keV, which might help to 
suppress the direct Urca process, in accordance with recent results from 
compact star cooling phenomenology. 
Nevertheless, since our results are strongly dependent on the parameters used 
and on the form of the regularization, more systematic studies are needed in
order to decide whether the CSL phase could be applied in the description of 
compact stars. 

\section*{Acknowledgements}

We are grateful to  J. Berdermann, M. Buballa,  H. Grigorian, 
G. R\"opke, N. Scoccola
and D.N. Voskresensky for their comments and interest in our work.
We acknowledge discussions with D. Rischke, I. Shovkovy and Q. Wang 
during the meetings of the Virtual Institute
VH-VI-041 of the Helmholtz Association on
{\it Dense hadronic matter and QCD phase transition} and 
with the participants
at the International Workshop on ``The new physics of compact stars'' at ECT* 
Trento, Italy.
D.N.A. acknowledges support from Landesgraduiertenf\"orderung,
Mecklenburg-Vorpommern and  from DAAD-Antorchas collaboration project under 
grant number DE/04/27956. 
D.N.A thanks for the hospitality of the Tandar Laboratory, CNEA, Buenos Aires, 
where part of this work has been concluded.

\end{document}